\documentclass[12pt]{article}
\usepackage{latexsym}
\usepackage{graphicx}
\usepackage{amsfonts, amsmath,amssymb}

\numberwithin{equation}{section}
\newcommand\xv{{\mathbf {l}}}

\newcommand\qv{{\mathbf {q}}}
\newcommand\Pd{\tfrac{1}{2}}
\newcommand\e{{\mathrm{e}}}

\newcommand\be{\begin{equation}}
\newcommand\ee{\end{equation}}
\newcommand\bee{\begin{eqnarray}}
\newcommand\eee{\end{eqnarray}}

\begin{document}
 \title{On the theory of spatially inhomogeneous Bose-Einstein condensation
of magnons in yttrium iron garnet
 }{}%
\author{{\it
 A.I.~Bugrij\footnote{E-mail: abugrij@bitp.kiev.ua},\addtocounter{footnote}{5}
 and V.M.~Loktev\footnote{E-mail: vloktev@bitp.kiev.ua} }\\
{\small\it N.N.~Bogolyubov Institute of Theoretical Physics of NAS of Ukraine, }\\
 {\small\it  Kiev 03143, Ukraine}
 }
 \date{}
 \maketitle

{\bf Abstract: } The Bose-Einstein condensation (BEC) of magnons created by a strong pumping in ferromagnetic
thin films of yttrium iron garnet used as systems of finite size is considered analytically. Such
a peculiarity, typical for this magnetic material, as the presence of a minimum in the spectrum of
spin waves at a finite value of the wave vector is taken into account. The definition of hightemperature
BEC is introduced and its characteristics are discussed. A role of boundary conditions
for spin variables is analyzed, and it is shown that in the case of free spins on the boundary
the magnon lattice can form in the system. The factors responsible for its appearance are discussed.


\section{Introduction}
Bose-Einstein condensation (BEC) is one of the few phenomena of
macroscopic physics, which have a quantum nature. A number of theoretical
and experimental studies are devoted to the study of BEC, and the history
is long and instructive (see, for example, the reviews Refs. 1--\,4 and
references therein). A remarkable feature of BEC is not so much the
possibility of its very existence as a phase transition of type II in a
Bose system, but the fact that it can take place in an ideal Bose gas or,
what is the same, in an ensemble of non-interacting particles or
quasiparticles. The latter are known to have their specificity. They are
related with excitations above the ground state of a many-body system, and
therefore characterized by a finite lifetime. Consequently, their BEC
should occur (and occurs) substantially in the non-equilibrium state,
which is also a subject of many studies (e.g., Refs. 5--11). This, in
turn, means that if a subsystem of quasiparticles, created by some
non-thermal way, usually obeying Bose-Einstein statistics
(excitons,$^{12-14}$ polaritons,$^{15-17}$ magnons,$^{5-11,18-21}$ photons
in a matter$^{22-24}$), is left to itself, then during establishing the
equilibrium one of the intermediate steps of relaxation may be the
formation of a Bose condensate. It exists and persists for a certain time,
during which one can talk about a finite value of the chemical potential
$\mu $ of corresponding quasiparticles. And if one does not make special
efforts to keep their number, the condensate as a collective of
(quasi-)particles with a density, which does not correspond to the
equilibrium one, will be damped, and the final step of evolution will be
the thermodynamic equilibrium (one can tell, ground) state of the system,
in which the number of excitations is determined only by a temperature $T$
when $\mu =0$.

The BEC, or the phenomenon of accumulation of Bose particles, an average
number $N$ of which persists that in particular for a gapless spectrum is
controlled by the condition $\mu <0$, in their lowest state was predicted
by Einstein many decades ago. According to modern concepts (see Ref. 25
and Refs. 1--\,4), the condensation corresponding to this phenomenon takes
place in a momentum (or energy) space, and there is no gas condensation in
a real coordinate space, and hence a condensed phase does not appear in
it. To be more precise, the spatial structure of the condensate reflects
only the coordinate distribution of the probability density of finding
particles in their ground state.

For observation of BEC in different systems many attempts were made, but
it was only recently realized experimentally in a particle
system.$^{26-28}$ The main obstacle in the realization of this phase
transition occurring when the chemical potential achieves the equality
$\mu (T_{\scriptscriptstyle{BEC}},N )=0$, is an extremely low temperature
$T_{\scriptscriptstyle{BEC}}$ at which the condensate starts to form and,
formally, the number of particles $N_0$ in the ground state tends to
infinity. Such a behavior is, of course, not physical, and it is generally
assumed that BEC corresponds to the condition $\mu
(T_{\scriptscriptstyle{BEC}},N )\to 0$; here $ N_0 \to N$, and the number
$N_{exc} =N-N_0 $ of other particles, or particles distributed over all
excited states is relatively small, $N_{exc} \ll N(\sim N_0 )$. As a
result, from the known formula$^{29}$
\begin{equation}
\label{eq1} T_{\scriptscriptstyle {BEC}} =3.31\frac{\hbar
^2}{k_{\scriptscriptstyle B} m}{\rm n}^{2/3},
\end{equation}
where $\hbar $ and $k_{\scriptscriptstyle B} $ are the Planck's and
Boltzmann's constants, $m$ is the mass of particles, and ${\rm n}$ is
their density. It is easy to see that for the rarefied gases of atoms of
alkali metals studied experimentally $T_{\scriptscriptstyle{BEC}}$ does
not exceed $10^{-6}-10^{-8}$ K. In the case of relatively light particles
(or quasiparticles), the situation seems to be more favorable, but also
for them (for example, for bosons with a mass of, say, electrons) the
temperature of the condensation $\sim $1--10 K, as follows from Eq.~(1.1),
is hard to achieve, requiring nearly limiting concentrations of
excitations in a crystal (see Ref. 12).

The greater and justified interest was attracted to the detection of
formation of a Bose-condensate of ferromagnons at almost room temperature
by means of Brillouin light scat\-te\-ring,$^{18}$ although the
low-temperature BEC of magnon gas of superfluid $^3$He was observed much
earlier.$^5$ Without denying this possibility, in principle, yet we note
that if BEC can actually occur at so high temperatures, then it (similar
to a high-temperature superconducti\-vity) can also be referred to as {\it
high-temperature}. But then there is the question about the causes or the
conditions under which such a physical phenomenon becomes realistic. And
the fact that it was possible to observe exactly for magnons is quite
natural. They, being typical of crystal elementary excitations, stand out
among others, first of all, by that they have a relatively long lifetime
due to the spin angular momentum conservation. Therefore, from the point
of view of studying quasi-stationary phenomena in which these excitations
are involved, or studying the behavior and measuring their characteristics
the magnons are more convenient to work with. The spin excitations in
superfluid $^3$He satisfy the same conditions (see Ref. 5).

Note two obvious reasons why the temperature of the condensation Eq.~(1.1)
for gases of quasiparticles is much higher than
$T_{\scriptscriptstyle{BEC}}$ of atomic gases. First, a mass of many
quasiparticles is much smaller than that of atoms: as is well known, even
charged carriers (electrons and holes) bound into neutral excitons in
semiconductors are lighter by an order of magnitude than free
electrons.$^{30,31}$ Second, in a gas of quasiparticles with a density of
$10^{18}-10^{20}$~cm$^{-3}$ can be achieved, which is much greater than
the density of atomic gases $10^{3}-10^{5}$~cm$^{-3}$, which are used in
corresponding BEC experiments. It is important that even for such a high
density the concentration of quasiparticles per cell is very small
$\sim10^{-5}-10^{-3}$, i.e., in a first approximation, their interaction
with each other can be neglected.

In this case, the fast enough spectral relaxation of a magnon
gas can be attributed to intense (but, what is significant,
persisting the number of magnons) exchange processes as
well as to relatively weak interactions with other objects,
quasiparticles of a different nature (e.g., phonons) and various
defects, including edges of a sample. In addition, the
magnon spectrum $\varepsilon ({\rm \tilde{{\bf q}}})$ (where ${\rm \tilde{{\bf
q}}}$ is the dimensional wave
vector), especially its gap $\varepsilon _0 $, can quite easily be controlled
by an external magnetic field ${\rm {\bf
H}}_0 $, which makes the investigation
of laws of their behavior at high densities even more
informative.
As shown in Ref. 21, the quasiparticles, the gap in the
spectrum of which and the temperature satisfy the inequality
\begin{equation}
\label{eq2}
k_B T\gg \varepsilon _0 ,
\end{equation}
reveal features of the transition from its initial nonequilibrium
distribution to equilibrium state. In particular, if
the lifetime permits, all non-equilibrium (i.e., pumped into
the system by one way or another) quasiparticles, not replenishing
the population of any of excited states, have time to
go to their lowest (but, again, which is also non-equilibrium
for the system as a whole) state via relaxation, which was
mentioned in Ref. 21 (see also Ref. 4). (The lifetime in principle
determines a threshold for observation of hightemperature
BEC, however not the phenomenon itself but
the presence of pumped magnons in the system. It is clear
that for a small lifetime magnons will have time to damp
before the pumping intensity provide the necessary increase
in their number$^4$). And it, in turn, is condensated by definition.
In this sense, the BEC of just these quasiparticles is carried
out at any (not only at extremely low, as is usually the
case) temperature and the whole system is always in the
BEC regime. In other words, the high-temperature BEC, or,
as noted, the accumulation of all non-thermal elementary
excitations brought in the system with the relevant behavior
of the chemical potential, indeed takes place for pumped
magnons.

In this regard, we note that for the above-written condition
which is satisfied with margin for magnons even in the
case of relatively low temperatures and high magnetic fields,
it should be taken into account that the total concentration ${\rm n}_0 $
of quasiparticles in their lowest state (as, in fact, in all
others) consists of two contributions, namely: ${\rm n}_0 ={\rm n}_0^{th}
+{\rm n}_0^{pump} $. The first of these, ${\rm n}_{0}^{th}$, is their thermal equilibrium
concentration in this state, and the second, ${\rm
n}_{0}^{pump}$, is its
growth after spectral redistribution of magnons appeared as
a result of strong external electromagnetic pumping $I_{pump}$.
Then, as is easy to check by direct calculation, the value of
${\rm n}_0 $ which depends, in fact, on the ratio $I_{pump}/(k_{\scriptscriptstyle{B}}T)$)$^{21}$ turns out
to be though much larger than in each of other quantum
states, but much less than in all excited states combined,
which, as known, is opposite to the situation typical of BEC
in its classical manifestation. Thus, we conclude that because
at high enough temperatures the inequality ${\rm n}_0 \ll {\rm n}_{exc} $ inevitably
remains in force, an assignment of “settling” of
non-equilibrium quasiparticles onto its ground level,
observed in Refs. 18 and 20, to the true BEC must be admitted,
to a certain extent, to be arbitrary. Moreover, the interpretation
of this phenomenon based on the formation of the
coherent collective state must also be performed with some
caution because the condensate in this situation can never be
quite intense, and its coherent properties require special examination.
Nevertheless, the noted special features of
exactly such a high-temperature BEC does not make it to be
a less interesting subject for theoretical and experimental
research.

In this paper, we aim to study the spatial (coordinate)
distribution of quasiparticles accumulating on its lowest
level (here, ferromagnons). The fact is that the experimental
measurements were carried out in thin ferromagnetic films
of yttrium-iron garnet (YIG) with an average size of about
$\sim$5~$\mu$m$\times$2~mm$\times$2~cm.$^{18,20}$ They are known not only for their
high quality, ensuring long lifetime of magnons even at high
densities, but also for the fact that the magnon spectrum in
these films is non-monotonic. Its minimum (see Ref. 32) is
located not at the least possible value of the wave vector, as
it is in the most of magnetic materials, but at some finite
value ${\rm
\tilde{{\bf q}}}_0 $, which is determined by the dipole-dipole interaction
between spins of iron ions, here $| {{\rm \tilde{{\bf q}}}_0 } |\approx
$3.4$\times$10$^{4}$~cm$^{-1}$.$^{20}$ This fact defines directly the space harmonic corresponding
to this lowest state, and inevitably leads to a non-monotonic
distribution of the quasiparticle density in a sample. We
emphasize that the case is an excited state of the system,
rather than a non-collinear spin-modulated structure of the
ground state of some magnetic materials, examples of which
are known.$^{33}$ In this case, periodic is not the spin direction in
the lattice, but the density of excitations. This manifests
itself in the formation in YIG films, pumped parametrically
by strong pulses of gigahertz range, of stripe periodic structures,
magnon lattices with a period of $\sim |{\rm \tilde{{\bf q}}}_0
|^{-1}$~cm, which
are nothing else than the equivalent of dynamic optical lattices,
emerging in particular as a result of self-diffraction in
experiments with coherent light beams.$^{34}$

In the case of magnons the situation is different, the process
of diffraction is absent, but there is a high-temperature
BEC with the formation of a standing wave to the intensity
of which, as will be seen, a significant contribution is made
by the thermal excitations. It is these magnon structures which
scatter test photons of the optical range. Then the
problem is not so much obvious and even quite trivial fact of
correspondence between the structure of the quasiparticle
spectrum and the spatial distributions of the probability densities
corresponding to it, but it is a search for the conditions
under which the stripe magnon structure can indeed survive
when there are superimposed contributions from the two
groups: the relatively weak, but supported by pumping, and
the strong thermal one. In addition, a rather unusual is the
fact that one of the critical factors of occurrence and observation
of the magnon lattice is a form of boundary conditions
that govern the spin variables at the sample boundaries (in a
real experiment, a thin ferromagnetic film).

\section{Model spectrum and general relations}
Let us assume that crystalline films of YIG, which were
investigated in experiments on BEC of magnons can be presented
as a parallelepiped with a volume ${\rm V}={\rm L}_x {\rm L}_y {\rm L}_z $ (Fig. 1),
and in general not only for ${\rm L}_x \ne {\rm L}_y \ne {\rm L}_z $ but for sufficiently
thin films ${\rm L}_x \ll {\rm L}_y \ll {\rm L}_z $. The number of sites $L_j (j=x,y,z)$
along each direction is related with the corresponding periods
$a_j $ of the lattice by usual relations: $L_j
={\rm L}_j /a_j +1$.

\begin{figure}[hpt]
\begin{center}
\includegraphics[height=60mm,keepaspectratio=true]
{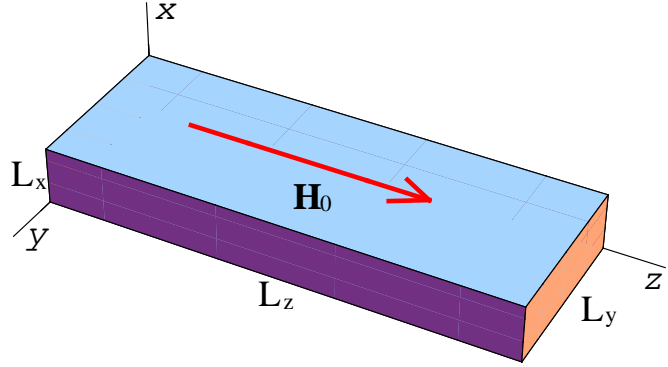} \caption {A schematic view of YIG film. The arrow indicates the direction of
magnetic field ${\rm{\mathbf{H}}}_{0}$ applied to it.}
\end{center} \end{figure}

As is known, in the case where the crystal has the ferromagnetic spin
ordering (for definiteness, the direction of the $z$-axis is the axis of
``easy'' magnetization, which, thus, will be the quantization axis), each
elementary excitation corresponds to one ``inverted'' spin in the site
${\rm {\bf l}}$.$^{35,36}$ (Strictly speaking, for an arbitrary spin $S$
of a paramagnetic ion, the number of such excitations (spin levels) on the
site can reach $2S$, but for further calculations such an increase is not
significant since we restrict ourselves to the so-called ``onemagnon''
approximation, i.e., we assume that there is no more than one excitation
in the site). From linear combinations of these states using known rules
it is easy to construct eigenstates of the multi-site translationally
invariant system in a form of spin waves with amplitudes $\psi _{\rm {\bf
q}} ({\rm {\bf l}})$, a direct form of which depends on boundary
conditions. For simplification (see Ref. 36) the periodic boundary
conditions are usually used, when the amplitudes are
\begin{equation}
\label{eq3} \psi _{\rm {\bf q}} ({\rm {\bf l}})=\prod\limits_{j=x,y,z}
{L_j }^{-\frac{1}{2}} e^{iq_j l_j }, \quad q_j =\frac{2\pi k_j}{L_j},
\quad k_j =0,1,...,L_j -1,
\end{equation}
i.e., have a form of plane waves, in which the dimensionless vector ${\rm
{\bf l}}$ enumerates the lattice sites: $l_j =1,{\kern 1pt}\,2,...,L_j $,
and the dimensionless wave vector $q_{j}=\tilde{q}_{j}a_{j}$ runs through
a discrete set with a step $\Delta q_j =2\pi /L_j $ within the first
Brillouin zone ($0 \leq q_j < 2\pi )$. In the case of free spins on the
boundary the solution of the problem with appropriate boundary conditions
gives somewhat other values for the amplitudes, namely (cf. Eq.~(2.1))
\begin{equation}
\label{eq4} \psi _{\rm {\bf q}} ({\rm {\bf l}})=\prod\limits_{j=x,y,z}
{L}^{-\frac{1}{2}}_{j} \gamma(q_{j})\cos q_j (l_j -\Pd), \quad q_j
=\frac{\pi k_j}{L_j},
\end{equation}
\[
\quad k_j =0,1,...,L_j -1, \quad \gamma(0)=1, \quad
\gamma(q\neq0)=\sqrt{2}.
\]
(It is interesting to note that the wave function has this form
only in the case of magnons, and if we consider, for example,
particles moving in a lattice with the conditions on the
boundary fixed for amplitudes, then in this case
$\psi _{\rm {\bf q}} ({\rm {\bf l}})\sim
\prod\limits_j {\sin q_j l_j } $). They, as is easy to verify (see the
Appendix), define eigenfunctions of the arising problem on
the eigenvalues of the Hamiltonian of the system and define
the full set of standing waves. The Brillouin zone is defined
somewhat differently: $0\leq q_j <\pi $, and the discrete step
$\Delta q_j =\pi/L_j$.

Whatever the boundary conditions for spins, they have
little effect on the spectrum of ferromagnons determined
mainly by the strong exchange interaction $J$, which for simplicity
is assumed to be isotropic. In this case, a kinetic component
of the spectrum can be represented by expression
valid for the case of both types of the boundary conditions
\begin{equation}
\label{eq5} \varepsilon _{kin} ({\rm {\bf q}})=4J\sum\limits_{j=x,y,z}
{\sin ^2\frac{q_j }{2}} .
\end{equation}
In the long-wavelength region $q_j \ll \pi$ the spectrum, as follows from
Eq.~(2.3) takes the usual ``quadratic'' form:
\begin{equation}
\label{eq6} \varepsilon _{kin} ({\rm {\bf q}})=J{\rm {\bf q}}^2,
\end{equation}
and it is easy to write it through the length of edges ${\rm L}_j $, periods
$a_j $ of the crystal lattice and the effective mass $m_{\scriptscriptstyle{M}} $ of
magnons directly related to the exchange integral $J$.

The written expression (2.4) does not account for the
gap, the value of which in ferromagnetic materials can be of
a dual nature. Without a magnetic field it is determined by
the magnetic anisotropy and is generally small. For the gap
in the case of YIG, more important is an external magnetic
field which is applied along the magnetization vector and
provides a variation in a fairly wide range of the quantity
$\varepsilon _0 =\mu _{\scriptscriptstyle{B}} gH_0 $
($\mu_{\scriptscriptstyle{B}}$ is the Bohr magneton, and $g$ is the $g$-factor of
three-valent iron ion close to two), taking into account which
the magnon spectrum takes the final form
\begin{equation}
\label{eq7}
\varepsilon ({\rm {\bf q}})=\varepsilon _0 +\varepsilon _{kin} ({\rm {\bf
q}}).
\end{equation}
The minimum values of the kinetic energy of magnons, following
from Eqs. (2.4) and (2.5), are provided by the smallest
wave vectors from the range enabled for them. In YIG
films, as already mentioned, it is not the case, and in the
direction of magnetization, or in our case, the quantization
axis $z$, due to the contribution of the same anisotropic interactions
the minimum is reached at some finite value of the $z$-projection
of the wave vector ${\rm {\bf q}}$: $q_{z} = q_0 $. An accurate albeit cumbersome expression
for the magnon dispersion in YIG
with taking into account its peculiar downward is known
(see Ref. 32), but for the problem of BEC it is sufficient also
to restrict to the quadratic approximation for the anisotropic
component. Then the kinetic energy of magnons can be
expressed as$^{4,20}$
\begin{equation}
\label{eq8} \varepsilon _{kin} ({\rm {\bf q}})=J\left[ {{\rm {\bf q}}_\bot
^2 +\eta(q_z )(q_z -q_0 )^2} \right].
\end{equation}
In what follows, the dimensionless coefficient $\eta(q_z )$ will be
assumed to be equal to unity without loss of generality.
Assuming an ideality of long-wavelength magnons (recall
that their interaction with each other $\sim ({\rm {\bf q}}_1 {\rm {\bf q}}_2 )^2$ (Ref. 30), and,
thus, is not large), we need to calculate the partition function
$Z$ of the grand canonical ensemble, which has the form
\begin{equation}
\label{eq9} \ln Z  =-\sum\limits_{\rm {\bf q}} \ln \left\{
1-\e^{[\mu-\varepsilon ({\rm {\bf q}})]/T}\right\},
\end{equation}
where $\varepsilon ({\rm {\bf q}})$ is the magnon energy defined in
Eq.~(2.5), with the wave number ${\rm {\bf q}}=(q_x ,q_y ,q_z )$, and $\mu
$ is the chemical potential of magnons. In the expression (2.7) and below
we use the system of units $\hbar =k_B =1$, restoring the dependence on
the fundamental values only where it is necessary. Then for the average
number of Bose-particles (the occupation number) in a quantum state with a
given wave vector, we have
\begin{equation}
\label{eq10} n_{\rm {\bf q}}=\frac{1}{\e^{[\varepsilon ({\rm {\bf
q}})-\mu]/T}-1}\,.
\end{equation}
As follows from Eqs. (2.7) and (2.8), the range of the value
of $\mu$ is limited by a minimum value of the energy. In the case
of the dispersion law (2.5) in the presence of the gap we
come to the inequality $\mu<\varepsilon _0 $, where $\varepsilon _0 = \varepsilon ({\rm {\bf
q}}_0 )$, and
${\rm {\bf q}}_0 =(0,0,q_0)$, and $\varepsilon _0 $ is completely determined by an external
static field. But if $\mu\to\varepsilon _0$, then $n_{{\rm {\bf q}}_0} \to \infty $ in a finite volume,
which is not possible.

To avoid this contradiction, we note that the chemical potential as an
independent thermodynamic variable is a convenient parameter in the
theory, but quite formal quantity if an experiment is considered where
$\mu $, as a rule, can be found only indirectly (for example, by
calculating it from the measured average number of particles or from other
observations). It is easy, however, to make sure that in studying the
ideal Bose gas, the chemical potential can be completely eliminated from
thermodynamic formulas, by replacing it with another independent variable
that has a clear physical meaning. If it is a BEC, a quite justified and
convenient quantity is the number of particles on the lowest level $n_0 $.
Indeed, using Eqs.~(2.5) and (2.8), we introduce the thermodynamic
variable
\begin{equation}
\label{eq11} n_0 =\frac{1}{\e^{(\varepsilon_{0}-\mu)/T}-1},
\end{equation}
which makes it possible to find not only
$\mu   =\varepsilon _0 -T\ln
(1+n_0^{-1})$, but all the others (for ${\rm {\bf q}}\ne {\rm {\bf q}}_0
)$) occupation
numbers
\begin{equation}
\label{eq11a} n_{{\rm {\bf q}}}=\frac{1}{(1+n_0^{-1})\e^{[
\varepsilon({\rm {\bf q}})-\varepsilon_0]/T} -1}= \frac{n_{\rm {\bf
q}}^{sat}n_{0}} {n_{{\rm {\bf q}}}^{sat}+n_{0} +1}\,.
\end{equation}
In Eq.~(2.10) we use the notation for the number
\begin{equation}
\label{eq12} n_{\rm {\bf q}}^{sat}=\frac{1}{\e^{\varepsilon _{kin} ({\rm
{\bf q}})/T}-1} \,,
\end{equation}
which is independent neither of the magnon spectrum gap in the magnon
spectrum defined by an external magnetic field nor of their chemical
potential and, as can be seen, which determines the maximum (saturating)
capacity of the state relative to accumulation in it of Bose particles at
a given temperature. The replacement, Eq.~(2.9), allows to avoid the
non-physical asymptotic $n_0 \to \infty $ and to study the most
interesting for BEC situation when $\mu   \to \varepsilon _0 $. The
parameterization of unknown quantities by the number $n_0 $ is convenient
also because in fact the condensate is identified with it, or
$N_{\scriptscriptstyle{BEC}} =n_0 $, and in addition, there is a
possibility to go correctly to the thermodynamic limit when it is
necessary.

Taking the condition $n_0 \gg n_{\rm {\bf q}}^{sat}$, corresponding to the
condensation regime, from Eq.~(2.10) for all ${\rm {\bf q}}\neq{\rm {\bf
q}}_{0}$ we obtain the expansion
\begin{equation}
\label{eq13} n_{\rm {\bf q}} =n_{\rm {\bf q}}^{sat}\left[ 1-\left(
\frac{n_{\rm {\bf q}}^{sat}+1}{n_0 }\right)+\left(\frac{n_{\rm {\bf
q}}^{sat}+1}{n_0 }\right)^2-...\right],
\end{equation}
which indicates that the value Eq.~(2.11) is indeed the maximum possible
(for the formal condition $n_0 \to \infty )$) occupation number of the
${\rm {\bf q}}$-state. The expansion (2.12) can also be used in writing
other thermodynamic quantities. Thus, the average number of particles
becomes
\begin{equation}
\label{eq14} N=\sum\limits_{{\rm {\bf q}}} n_{\rm q}=n_0
+\sum\limits_{{\rm {\bf q}}\ne {\rm {\bf q}}_0 }  n_{\rm q}=n_0 +N_{exc},
\end{equation}
where the quantity
\begin{equation}
\label{eq15} N_{exc} =\sum\limits_{{\rm {\bf q}}\ne {\rm {\bf q}}_0}
n_{\rm {\bf q}}^{sat}\left[ 1-\left( \frac{n_{\rm {\bf q}}^{sat}+1}{n_0
}\right)+\left(\frac{n_{\rm {\bf q}}^{sat}+1}{n_0
}\right)^2-...\right]\approx N_{sat} -\frac{\delta N_{sat}}{n_0 }
\end{equation}
for a given $n_0 $ defines the total density of magnons in all
excited states. From it in the last relation we separated out
the number
\begin{equation}
\label{eq16} N_{sat}=\sum\limits_{{\rm {\bf q}}\ne {\rm {\bf q}}_0 }
  n_{\rm {\bf q}}^{sat},
\end{equation}
which specifies the maximum, reached at the same (see
above) condition $n_0 \to \infty$, number of thermal excitations,
and the coefficient
\begin{equation}
\label{eq17} \delta N_{sat}=\sum\limits_{{\rm {\bf q}}\ne {\rm {\bf q}}_0
}n_{\rm {\bf q}}^{sat}(n_{\rm {\bf q}}^{sat}+1)
\end{equation}
for the first in $1/n_0 $ correction determining thermal fluctuations of
the quantity Eq.~(2.15). In this case, it is seen that regardless of the
specific type (type of Eq.~(2.6)) of the dispersion of elementary
excitations the both quantities $N_{sat}$ and $\delta N_{sat}$   are
monotonically increasing functions of temperature, since with increasing
$T$ all occupation numbers $n_{\rm {\bf q}}^{sat}$ are also growing.

Finally, combining Eqs.~(2.13), (2.15), and (2.16) in the natural
assumption $N\gg 1$ (and, of course, $n_0 \gg 1)$), we arrive at the
equation
\begin{equation}
\label{eq18} N\simeq n_0 +N_{sat} -\frac{\delta N_{sat}}{n_0 }
\end{equation}
for finding the density of a Bose condensate. In particular, from
Eq.~(2.17) it follows that the critical temperature of BEC is a solution
of the equation $N_{sat} (T_{\scriptscriptstyle{BEC}} )=N$ and already
depends on both the total concentration of particles and the specific form
of the spectrum. And an explicit expression for the temperature
$T_{\scriptscriptstyle{BEC}} $ is also defined by the dimensionality and
even the shape of a sample, including the conditions on all its
boundaries.

Similar quantities can be introduced for thermal magnons
when $\mu =0$, separating in the same way the equilibrium
population
\begin{equation}
\label{eq19} n_0^{th} =\frac{1}{\e^{\varepsilon _0 /T} -1}
\end{equation}
and representing, respectively,
\begin{equation}
\label{eq20} n_{\rm {\bf q}}^{th} =\frac{n_{\rm {\bf q}}n_{0}^{th}
}{n_{\rm {\bf q}}+n_{0}^{th} +1},
\end{equation}
where $n_{\rm {\bf q}} $ is the same occupation number Eq.~(2.12) as in
Eq.~(2.10). Remembering now that under pumping all the populations can be
divided into two components: $n_{\rm {\bf q}}=n^{th}_{\rm {\bf q}}+n_{\rm
{\bf q}}^{pump}$, with the help of Eqs.~(2.14) and (2.19) we find that for
all ${\rm {\bf q}}\ne {\rm {\bf q}}_0 $
\begin{equation}
\label{eq21} n_{\rm {\bf q}}^{pump} \approx \frac{\mu}{T}n_{\rm {\bf q}}
(n_{\rm {\bf q}} +1).
\end{equation}
In other words, because of the condition $\mu   \ll T$, the populations
of excited states are little changed, while the density of
the pumped particles in a Bose condensate behaves completely
different
\[
n_0^{pump} \approx \frac{T}{\mu _{\scriptscriptstyle B} gH_0 }\frac{\mu
}{(\mu _{\scriptscriptstyle B} gH_0 -\mu )},
\]
or can be arbitrarily large when approaching $\mu   \to \mu
_{\scriptscriptstyle B} gH_0 $. Moreover, from the last expression it
follows that for the condensate state (in contrast to Eq.~(2.20)), on the
contrary, there is even an ``enhancement factor'' because $T\gg \mu
_{\scriptscriptstyle B} gH_0 $. To some extent, the increase due to the
growth of the total number of excitations of the quantity $n_{0}$ only is
consistent with the description of the magnon BEC based on the
phenomenological approach, in which the dynamic macroscopic magnetization,
which is identified with the condensate, is entirely caused by
pumping$^{10}$ (see also Ref.~4).

As a result we come to the conclusion mentioned already
earlier: Magnons created in a thermally populated
ferro-system by an intense electromagnetic pump are accumulates
mainly on its lowest level. Consequently, such a
subsystem of quasiparticles artificially created at any temperature,
satisfying the inequality $T\gg \varepsilon_{0}$ (1.2), by itself formally
(if the lifetime allows) undergoes BEC, coexisting
with a powerful thermal collective of particles identical to it.
It is this BEC which is essentially high-temperature, and its
accompanying process is nothing like a turning of the magnetic
crystal with the anomalously excited lowest mode into
the magnon tuning fork.

\section{A spatial distribution of density of excitations}
We define the single-particle distribution function of
magnons as follows:
\begin{equation}
\label{eq22} \rho ({\rm {\bf l}})=\sum\limits_{\rm {\bf q}} {n_{\rm {\bf
q}}} \left| {\psi _{\rm {\bf q}} ({\rm {\bf l}})} \right|^2,
\end{equation}
where $\psi _{\rm {\bf q}}({\rm {\bf l}})$ is the amplitude of a spin
wave. In the case of periodic boundary conditions (2.1) $|\psi_{\rm {\bf
q}} ({\rm {\bf l}})|^2=(L_{x}L_{y}L_{z})^{-1}$ and the distribution
function Eq.~(3.1) does not depend on ${\rm {\bf l}}$. For free boundary
conditions the amplitude is real and expressed in the form of a ``standing
wave'', Eq.~(2.1). Therefore, separating in the sum Eq.~(3.1) the term
with the highest value of $n_{\rm {\bf q}}$, namely at ${\rm {\bf q}}={\rm
{\bf q}}_{0}$, according to Eq.~(2.6), we obtain
\begin{equation} \label{eq23} \rho
({\rm {\bf
l}})=\frac{2n_{0}}{L_{x}L_{y}L_{z}}\cos^{2}q_{0}(l_{z}-\Pd)+\sum\limits_{{\rm
{\bf q}}\neq {\rm {\bf q}}_{0}} {n_{\rm {\bf q}}}  {\psi^{2} _{\rm {\bf
q}} ({\rm {\bf l}})}.
\end{equation}
With increasing $n_{0}$ the second term in Eq.~(3.2) reaches the limit
$$\rho_{sat} ({\rm {\bf
l}})=\sum\limits_{{\rm {\bf q}}\neq {\rm {\bf q}}_{0}} {n^{sat}_{\rm {\bf
q}}}  {\psi^{2} _{\rm {\bf q}} ({\rm {\bf l}})}, $$ So, sooner or later,
the first term in Eq.~(3.2) becomes dominant. As a result, $\rho({\rm {\bf
l}})$ oscillates with a frequency $q_{0}$ along the $z$-axis. Thus,
qualitatively the mechanism of the appearance of the periodic structure in
the spatial density of magnons turns out to be relatively simple. However,
an exact calculation of the function Eq.~(3.2) is rather cumbersome, so it
makes sense to consider first the one-dimensional case. This will allow to
understand how and under what conditions the lattice can be formed in the
pumped system of magnons at temperatures exceeding the spectral gap.

\bigskip
\noindent{\bf 3.1. A one-dimensional ferromagnet.}

\medskip
In the case of free boundary conditions the wave function and the
quasi-momentum are written as follows (see the Appendix): \be\psi _{q} (l
)=L^{-\frac{1}{2}}\gamma(q)\cos q (l -\Pd), \quad q =\frac{\pi k }{L },
\quad k =0,1,...,L -1\,. \label{eq23a}\ee The distribution Eq.~(3.1) is
also easily rewritten and takes the form
\begin{equation}
\label{eq24} \rho(l)=\sum\limits_{q } {n_{q}}  {\psi^{2} _{q } (l )},
\end{equation}
in which
\begin{equation}
\label{eq25} n_{q}= \frac{n_{q }^{sat}n_{0} }{n_{q }^{sat}+n_{0} +1},
\quad n_{q }^{sat} =\frac{1}{\e^{\varepsilon _{kin} (q )/T} -1}\,.
\end{equation}

Consider first the ``monotonous'' dispersion
$\varepsilon(q)=\varepsilon_{0}+J q^{2}$. Because the quasi-momentum $q $
is a discrete quantity (see Eq.~(3.3)), instead of the occupation number
$n_{q}$ we introduce the function of an integer argument
\begin{equation}
\label{eq26} f(k)=n_{q}=\frac{1} {e^{\delta +\omega ^2k^2}-1}\, , \quad
\delta =\ln (1+n_0^{-1} ), \quad \omega ^2=\frac{J}{\pi^2L^2 T}.
\end{equation}
Then the spatial density of magnons Eq.~(3.4) for the case Eq.~(2.10) can
be written as the sum
\begin{equation}
\label{eq27} \rho (z)=\frac{1}{L}\sum\limits_{k=0}^\infty
{f(k)\gamma^{2}(k)\cos ^2\pi kz=\frac{1}{2L}\left[ {\rho _1 (z)+\rho _1
(0)} \right]} ,
\end{equation}
where the variable $z=(l-1/2)/L$ varies in the range
$0<z<1$. $\rho_{1}(z)$ denotes the sum
\begin{equation}
\label{eq28} \rho _1 (z)=\sum\limits_{k=-\infty }^{\infty } {f(k)\cos 2\pi
kz}.
\end{equation}
Note that from the definition Eq.~(3.7) it follows that $\rho
(z)=\rho(1-z)$, i.e., the magnon density is a symmetric function with
respect to the point $z=1/2$.

Considering the variable $k$ as complex, we express the desired sum
Eq.~(3.8) via the contour integral
\begin{equation}
\label{eq29} \rho _1  (z)=\oint\limits_{C_1 }\frac{{\rm d}k\, \e^{2\pi
ikz}}{\e^{2\pi ik}-1}\,f(k) \, ,
\end{equation}
where the integration path $C_1 $ is shown in Fig. 2.
\begin{figure}[h]
\begin{center}
\includegraphics[height=70mm,keepaspectratio=true]
{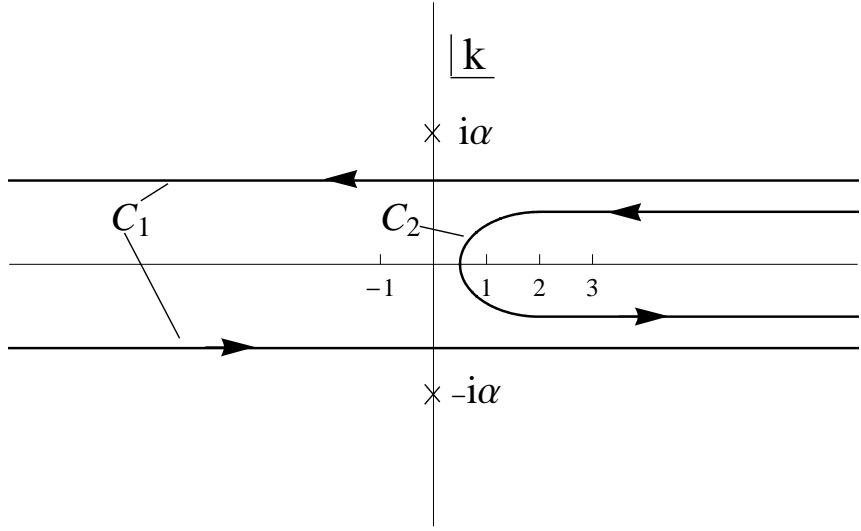} \caption {$C_{1}$, $C_{2}$ -- are the contours of integration
in the integrals Eqs.~(3.9) and (3.17). The crosses indicate the closest
to the real axis of the complex $k$-plane poles of the functions $f(k)$,
Eq.~(3.6), and $F(k)$, Eq.~(3.31).}
\end{center} \end{figure}

\noindent Since, as is easily seen, the function Eq.~(3.6) has simple
poles at $k=k_{j}$, where
\begin{equation}
\label{eq30} \omega k_j =\pm iu_j \pm \frac{\pi j}{u_j} , \quad u_j =\sqrt
{\frac{\delta }{2}+\sqrt {\frac{\delta ^2}{4}+\pi^{2} j^2} } , \quad
j=0,1,...\quad ,
\end{equation}
the integral Eq.~(3.9) is expressed by the sum of residues at these poles,
in which only contributions from the poles closest to the real axis in the
complex $k$-plane are taken into account. After having done the necessary
calculations, we obtain the expression
\begin{equation}
\label{eq31} \rho _1  (z)= {\frac{\pi^{2} }{\omega^{2} }} \frac{\cosh
\alpha (1-2z)}{\alpha \cosh \alpha }+o[ \e ^{-\pi^{3/2}z/\omega}], \quad
\alpha =\frac{\pi }{\omega }\sqrt{\delta}\, ,
\end{equation}
included in the formula Eq.~(3.7) for $\rho  (z)$. It, however, contains
the quantity $\rho _1  (0)$, which does not follow from Eq.~(3.11) because
of the large value of corrections to $z=0$. To calculate $\rho _1(z)$ in
the range of small values of $z$ it is necessary to replace Eq.~(3.9) with
the representation \bee\nonumber &&\rho_1
(z)=-\frac{i}{2}\oint\limits_{C_1} {\rm
d}k\,f(k)\cos 2\pi kz\, \cot\pi k=\\
&&=\frac{2\pi^{2}}{\omega^{2}\alpha} \frac{\cosh 2\alpha z}{\e^{2\alpha}
-1}+\int\limits_0^\infty {\rm d}k\,f(k)\cos 2\pi kz+o[\e^{-\pi
^{3/2}/\omega} ] , \label{eq32a}\eee
from where we find
\begin{equation}
\label{eq32} \rho_1(0)= \frac{2\pi^{2} }{\omega^{2}\alpha(\e^{2\alpha} -1)
} +\frac{\sqrt \pi }{2\omega }Li_{\frac{1}{2}} (e^{-\delta })=n_0 \alpha
\cosh\alpha +\frac{\sqrt \pi }{\omega }\left[ {Li_{\frac{1}{2}}
(e^{-\delta })-\sqrt {\frac{\pi }{\delta }} } \right],
\end{equation}
where $Li_{\frac{1}{2}} (e^{-\delta })$ -- is the polylogarithm.$^{37}$ The found dependencies
solve the problem of the spatial distribution of magnons,
but in the assumption of a quadratic spectrum it is
smooth not only for the case of periodic boundary conditions,
but also in the case of free boundaries.

The situation is different if the spectrum has a minimum
at ${q}={q}_0 \ne 0$, which in the one-dimensional case can be represented
without loss of generality as
\begin{equation}
\label{eq33} \varepsilon (q )=\varepsilon _0 +J(q -q_0 )^2.
\end{equation}
Then, using Eq.~(3.14), the expression (3.4) takes the form \bee\nonumber
&&\rho (z)=\frac{1}{L}\sum\limits_{k=0}^\infty f(k-k_0
)\gamma^{2}(k)\cos ^2\pi kz=\frac{1}{L}[A_{latt}(z)\cos ^2\pi k_0 z+A(z)],\\
&&A_{latt}(z)=2\rho _1 (z),\quad A(z)=\rho _1 (0)-\rho _1 (z)-\rho _2
(0)-\rho _2 (z)-f(k_{0}),\label{eq34a}\eee in which we separated out the
term oscillating with a characteristic period $\sim k_0^{-1} =\pi /(Lq_0)$
completely caused by the presence of cosine in the wave function
Eq.~(3.3). The function $\rho _1  (z)$ in Eq.~(3.15) is defined in
Eq.~(3.8), and the function
\begin{equation}
\label{eq34} \rho _2  (z)=\sum\limits_{k=1}^\infty {f(k+k_0 )\cos 2\pi kz}
\end{equation}
can also be represented by the contour integral
\begin{equation}
\label{eq35} \rho _2  (z)=\oint\limits_{C_2 } \frac{{\rm d}k\,\e^{2\pi
ikz} }{\e^{2\pi ik}-1}\,f(k+k_0 )\, ,
\end{equation}
where the contour of integration $C_{2}$ is shown in Fig.~2. A direct
calculation of this integral as well as Eq.~(3.9) we perform in two steps.
First, we assume that the aboveintroduced parameters $\delta $ and $\omega
$ are small, and the value of $k_0 $ is high that corresponds to the
situation which occurs in study of YIG. Then, using the exact value of the
integral
\[
\int\limits_{-\infty }^\infty {\frac{{\rm d}\xi\,\e^{2\pi \xi z}
}{\e^{2\pi \xi} +1} } =\frac{1}{2\sin \pi z}
\]
and expanding in a series the function $f(k+k_0 )$ in the neighborhood of
the point $k=1/2$, we find the asymptotic expansion of the integral
Eq.~(3.17). In the first order we have
\begin{equation}
\label{eq36} \rho _2  (z)\simeq-f_0 \,{\rm Re}\left( \frac{\e^{-\kappa}
}{1-\e^{-2\kappa-2\pi i z} } \right),
\end{equation}
where $f_0 = f(k_0 +1/2)$, $\kappa = f_1 /2f_0 $ and $f_1 = f'(k_0+1/2)$. The
expression (3.18) is a good approximation in the range of
$k_0^{-1}<z<1-k_0^{-1} $, and at lower values of $z$ it is necessary to
use the representation
\bee \label{eq36a} &&\rho _2
(z)=-\frac{i}{2}\oint\limits_{C_2 } {\rm d}k\, f(k+k_0
)\cos 2\pi k z\,\cot\pi k = I_1 (z)+I_2 (z),\\
\nonumber &&I_1 (z)=2{\rm Re}\int\limits_{C_- } \frac{{\rm d}k\,\cos 2\pi
k z}{\e^{2\pi i k} -1}\, f(k+k_0 ) , \quad I_2
(z)=\int\limits_{1/2}^\infty {\rm d}k\,f(k+k_0 )\cos 2\pi k z , \eee in
which $C_- $ denotes a part of the contour $C_2 $, which lies in the lower
half-plane of the complex variable $k$. An approximate expression for the
first integral Eq.~(3.19) can be obtained by analogy with Eq.~(3.18) which
gives
\begin{equation}
\label{eq37} I_1 (z)\simeq -\frac{1}{24}(2\pi zf_0 \sin \pi z-f_1 \cos \pi
z).
\end{equation}
The second integral can also be found after simple calculations.
We write it only for the point $z=0$
 \bee\nonumber &&I_2
(0)=\frac{1}{\omega \sqrt \delta }\,\arctan\left[\frac{\sqrt \delta
}{\omega (k_0 +1/2)}\right]+\frac{\sqrt \pi }{2\omega }Li_{\frac{1}{2}}
(e^{-\delta })-\frac{\pi
}{2\omega \sqrt \delta }+\\
&& +\int\limits_0^{k_0 +1/2}{\rm d}k {\left( {\frac{1}{\delta +\omega
^2k^2}-\frac{1}{\e^{\delta +\omega ^2k^2} -1}} \right)}\,
.\label{eq37a}\eee

As a result, from Eqs.~(3.18)--(3.21) it is seen that even at a small
distance from the edges of the chain between the two contributions $\rho
_2  (z)$ and $\rho _2  (0)$ there is the inequality $\rho _2  (0)\gg \rho
_2  (z)$, and only in the immediate vicinity of the boundaries, these
contributions become comparable. Then for the amplitude Eq.~(3.15) of the
periodic structure using Eq.~(3.11) we find
\begin{equation}
\label{eq39} A_{latt} (z)=2n_0 \alpha\, \coth\alpha -\frac{4\pi^{2}
}{\omega^{2} } \lambda (z), \quad \lambda (z)=\frac{\sinh(\alpha
z)\,{\kern 1pt}\sinh\alpha (1-z)}{\alpha \sinh\alpha }.
\end{equation}
The expression for the non-uniform and “not-lattice” distribution $A(z)$
in Eq.~(3.15) has a more complicated form, but after calculating the
integrals Eqs.~(3.17) and (3.19) with Eqs.~(3.20) and (3.21) one can
obtain fairly simple approximate formula for it
\begin{equation}
\label{eq40} A(z)=\frac{2}{\omega ^2}\left[ {\pi ^2\lambda (z)-O\left(
{\frac{1}{2k_0 +1}} \right)} \right].
\end{equation}

Using the explicit form of the functions $A_{latt} (z)$ and $A(z)$, which
enters the distribution of the magnon density in the chain Eq.~(3.15), one
can find the conditions for the formation of the magnon lattice. The
corresponding periodic structure arises in the case when its amplitude
$A_{latt} (z)$ exceeds the value of the relatively smooth contribution
$A(z)$, which is physically controlled by the ratio $a_{latt} (z)=A_{latt}
(z)/A(z)$. For small $n_0 $ of the condensate component of Bose particles,
or if the inequalities $\omega ^2n_0 \ll 1$, $\alpha \gg 1$ are satisfied,
the coefficient $a_{latt}(z)\approx \Pd\e^{-2\alpha z} \ll 1$, so that the
oscillations cannot appear as, in fact, expected. Moreover, there is no
periodic structure in equilibrium, although the population $n_0^{th} $
(see Eq.~(2.18)) is the greatest. On the contrary, in the case of large
numbers $n_0 $ and the small gap in the magnon spectrum when $\alpha \ll
1$, the quantity $a_{latt}(z)\approx n_0 \omega ^2/[\pi ^2z(1-z)]\gg 1$,
and the lattice of the excitation density dominates over the background.
In this case, the critical (transition) density of a Bose condensate, as
is easily seen, is its value at which $a_{latt} (z)\approx 1$, or $n_0
\approx 1/\omega ^2=\pi ^2L^2 T/J$. It should also be noted that the
appearance of the modulated structure at a relatively smooth background of
the density distribution of pumped excitations takes place with increasing
the pumping (i.e., the density of the Bose condensate) gradually, not
abruptly, as stated in Ref.~9, and does not require the assumption about
the presence in the system of two condensates$^{25}$ or an abnormally
strong damping of quasiparticles (magnons).

\bigskip
\noindent {\bf 3.2. A three-dimensional ferromagnet.}

\medskip
We represent the ratio $\varepsilon_{kin}(\qv)/T$ in terms of
dimensionless parameters
 \be\label{7v}
\frac{\varepsilon_{kin}(\qv)}{T}=\omega_{x}^{2}k_{x}^{2}+\omega_{y}^{2}k_{y}^{2}+
\omega_{z}^{2}(k_{z}-k_{0})^{2},\ee
where $k_{0}=22282$ is the integer number closest to $\tilde{q}_{0}{\rm L}_{z}/\pi$ at
${\rm L}_{z}=2$~cm, and
  \be\label{8v}\omega_{x}=\frac{{\rm L}_{z}}{{\rm L}_{x}}\omega,
 \quad \omega_{y}=\frac{{\rm L}_{z}}{{\rm L}_{y}}\omega,\quad
 \omega_{z}=\omega=\frac{\pi}{{\rm L}_{z}}\sqrt{\frac{J}{T}}.\ee
 If the constant $J$ is expressed via the magnon mass $m_{mag}$ (in
YIG it is ${}\simeq5m_{e}$), we obtain, that in the room-temperature
range ($T\simeq300\,K$)
 \be\label{9v} \omega=\frac{\pi \hbar}{{\rm L}_{z}}
 (2m_{mag}k_{\scriptscriptstyle B}T)^{-\frac{1}{2}}\simeq8.53\times10^{-8}.\ee

 For a quantitative description of the effect of oscillations
it is necessary to calculate the sum in rhs of Eq.~(3.2). But instead it
is easier to consider the magnon density averaged over the coordinates
$\xv_{\perp}$, namely
 \be\label{11v}
\bar{\rho}(z)=L_{z}\sum_{\xv_{\perp}}\rho(\xv)=\sum_{\qv}n_{\qv}\gamma^{2}(q_{z})
\cos^{2}q_{z}(l_{z}-\Pd)
=\sum_{k_{z}=0}^{\infty}F(k_{z}-k_{0})\gamma^{2}(k_{z})\cos^{2}\pi k_{z}
z,\ee where \be\label{12v} F(k_{z}-k_{0})=\sum_{\qv_{\perp}}n_{\qv}, \quad
z=\frac{l_{z}-1/2}{L_{z}}.\ee Represent $\bar{\rho}(z)$, similar to the
one-dimensional case Eq.~(3.15), as:
 \bee\label{13v}
&&\bar{\rho}(z)=A_{latt}(z)\cos^{2}\pi k_{0}z+A(z),\\\nonumber
&&A_{latt}(z)=2\rho_{1}(z),\quad
A(z)=\rho_{1}(0)-\rho_{1}(z)-\rho_{2}(0)-\rho_{2}(z)-F(k_{0}),\eee where
\be\label{14v} \rho_{1}(z)=\sum_{k_{z}=-\infty}^{\infty}F(k_{z})\cos 2\pi
k_{z}z,\quad \rho_{2}(z)=\sum_{k_{z}=1}^{\infty}F(k_{z}+k_{0})\cos 2\pi
k_{z}z.\ee As can be seen from the definition Eq.~(3.28), the function
$F(k_{z})$ is a sum of “Bose”-terms of the type of Eq.~(3.6)
 \be\label{15v}
F(k_{z})=\sum_{k_{x}=0}^{\infty}\sum_{k_{y}=0}^{\infty}\frac{1}{\e^{w}-1},\ee
where \be\label{16v}
w=w(k_{x},k_{y},k_{z})=\delta+\omega_{x}^{2}k_{x}^{2}+\omega_{y}^{2}k_{y}^{2}+
\omega_{z}^{2}k_{z}^{2}.\ee Therefore, to calculate $\rho_{1}(z)$ one can
use the transformations Eq. (3.9) when $k_{0}^{-1}<z<1-k_{0}^{-1}$, of
Eq.~(3.12) when $0\leqslant z<k_{0}^{-1}$. As a result, we obtain for
$\rho_{1}(z)$ a series representation
 \be\label{17v}
\rho_{1}(z)=\frac{\pi^{2}}{\omega_{z}^{2}}\sum_{k_{x}=0}^{\infty}
\sum_{k_{y}=0}^{\infty}\frac{\coth\alpha_{z}(1-2z)}{\alpha_{z}\sinh\alpha_{z}},\ee
where $\alpha_{j}$ denote the functions \bee\nonumber
&&\alpha_{x}=\alpha_{x}(k_{y},k_{z})=\frac{\pi}{\omega_{x}}\sqrt{\delta+\omega_{y}^{2}k_{y}^{2}
+\omega_{z}^{2}k_{z}^{2}},\\\label{18v}
&&\alpha_{y}=\alpha_{y}(k_{x},k_{z})=\frac{\pi}{\omega_{y}}\sqrt{\delta+\omega_{x}^{2}k_{x}^{2}
+\omega_{z}^{2}k_{z}^{2}},\\\nonumber
&&\alpha_{z}=\alpha_{z}(k_{x},k_{y})=\frac{\pi}{\omega_{z}}\sqrt{\delta+\omega_{x}^{2}k_{x}^{2}
+\omega_{y}^{2}k_{y}^{2}}.\eee When $\delta\ll1$ and taking into account
the smallness of the ratio $\omega_{z}/\omega_{x}$,
$\omega_{z}/\omega_{y}$, for Eq.~(3.33) there is a valid approximate
expression \bee\nonumber &&\rho_{1}(z)=
\frac{\pi^{2}}{\omega_{z}^{2}}\,\frac{\coth\alpha(1-2z)}{\alpha\sinh\alpha}-
\frac{\pi}{\omega_{z}\omega_{y}}\ln(1-\e^{-2\pi z\omega_{y}/\omega_{z}})=\\
&&=\frac{\alpha\coth\alpha}{\delta}+\frac{2\pi^{2}}{\omega_{z}^{2}}\,\frac{\sinh\alpha
z\cdot
\sinh\alpha(1-z)}{\alpha\sinh\alpha}-\frac{\pi}{\omega_{z}\omega_{y}}
\ln(1-\e^{-2\pi z\omega_{y}/\omega_{z}}),\label{19v}\eee where
$\lambda(z)$ is defined in Eq.~(3.22) and $
\alpha=\alpha_{z}(0,0)={\pi}\sqrt{\delta}/{\omega_{z}}$. The discarded
contributions in Eq.~(3.35) does not exceed the value ${\pi}\e^{-2\pi
z\omega_{x}/\omega_{z}}/(2\omega_{x}\omega_{z})$. As before, for finding
the value of $\rho_{1}(0)$ we use the transformation Eq.~(3.12). Then
\be\label{21v} \rho_{1}(0)=\sum_{k_{z}=-\infty}^{\infty}F(k_{z})=R+I,\ee
where $R$ is the sum of residues in the closest to the real axis poles of
the function $F(k_{z})$ \be\label{22v}
R=\frac{2\pi^{2}}{\omega_{z}^{2}}\sum_{k_{x}=0}^{\infty}
\sum_{k_{y}=0}^{\infty}\frac{1}{\alpha_{z}(\e^{2\alpha_{z}}-1)}\,.\ee
Similarly to Eq.~(3.35) we get for $R$ the approximate expression
\be\label{23v}
R=\frac{2\pi^{2}}{\omega_{z}^{2}}\,\frac{1}{\alpha(\e^{2\alpha}-1)}=\frac{\alpha\coth\alpha}{\delta}-
\frac{\pi}{\omega_{z}}\,\frac{1}{\delta^{1/2}}\,,\ee and the neglected
terms do not exceed
${\pi}\e^{-2\pi\omega_{y}/\omega_{z}}/({\omega_{y}\omega_{z}})$. $I$ on
rhs of Eq.~(3.36) denotes the integral \be\label{24v}
I=2\int\limits_{0}^{\infty}{\rm d}k_{z}\, F(k_{z}).\ee We apply
consistently the transformation Eq.~(3.7) to the summations in Eq.~(3.31).
Then, in particular, for the sum over $k_{y}$ we obtain \be\label{25v}
\sum_{k_{y}=0}^{\infty}\frac{1}{\e^{w}-1}=\frac{\pi^{2}}{\omega_{y}^{2}}\,\frac{1}{\alpha_{y}}\,
\frac{1}{\e^{2\alpha_{y}}-1}+\frac{1}{2}
\int\limits_{-\infty}^{\infty}\frac{{\rm
d}k_{y}[1+\delta(k_{y})]}{\e^{w}-1}\,,\ee where $\delta(k)$ is the
$\delta$-function of Dirac. Performing similarly to Eq.~(3.40) the
summation over $k_{x}$, we obtain for $F(k_{z})$ the following expression:
\be\label{26v}
F(k_{z})=\frac{1}{4}\int\limits_{-\infty}^{\infty}\frac{{\rm d}k_{x}\,{\rm
d}k_{y}}{\e^{w}-1}[1+\delta(k_{x})][1+\delta(k_{y})]
+\frac{\pi^{2}}{2\omega_{x}^{2}}\int \limits_{-\infty}^{\infty}\frac{{\rm
d}k_{y}}{\alpha_{x}}\,\frac{1+\delta(k_{y})}{\e^{2\alpha_{x}}-1}+
\frac{\pi^{2}}{\omega_{y}^{2}}\sum_{k_{x}=0}^{\infty}\frac{1}{\alpha_{y}}\,\frac{1}
{\e^{2\alpha_{y}}-1}.\ee Note that in the sum on rhs of Eq.~(3.41) only
the term with $k_{x}=0$ can remain, since the other terms are
“exponentially” small: $\sim\e^{-2\pi\omega_{x}/\omega_{z}}$. Substituting
Eq.~(3.41) into Eq.~(3.39) and integrating, we obtain for the integral $I$
the following expression: \bee\nonumber
I&=&\frac{\pi^{3/2}}{8\omega_{x}\omega_{y}\omega_{z}}Li_{\frac{3}{2}}(\e^{-\delta})-
\frac{\pi(\omega_{y}+\omega_{x})}{8\omega_{x}\omega_{y}\omega_{z}}\ln(1-\e^{-\delta})
+\frac{\sqrt{\pi}}{8\omega_{z}}Li_{\frac{1}{2}}(\e^{-\delta})-\\
\label{27v}&&-\frac{\pi}{4\omega_{y}\omega_{z}}\ln(1-\e^{-2\pi\delta^{1/2}/\omega_{x}})+
\frac{\pi}{\omega_{y}\omega_{z}}G\biggl(\frac{2\pi}{\omega_{y}}\delta^{1/2}\biggr)+
\frac{\pi}{2\omega_{x}\omega_{z}}G\biggl(\frac{2\pi}{\omega_{x}}\delta^{1/2}\biggr)\,,\eee
where $G(t)$ denotes the integral \be\label{28v}
G(t)=\int\limits_{0}^{\infty}\frac{{\rm d}x}{\sqrt{t^{2}+x^{2}}}\,
\frac{1}{\e^{\sqrt{t^{2}+x^{2}}}-1}\,.\ee The asymptotic expansion of
$G(t)$ for $t\ll1$ is of the form \be\label{29v}
G(t)=\frac{\pi}{2t}+\frac{1}{2}\ln\frac{t}{4\pi}+{\Bbb C}
-\frac{\zeta(3)}{16\pi^{2}}t^{2}+O(t^{4}),\ee where ${\Bbb C}$ is the
Euler’s constant, $\zeta(\theta)$ is the $\zeta$-function of Riemann. The
calculation of $\rho_{2}(z)$ is not fundamentally different from the
one-dimensional case, Eq.~(3.18). It is only necessary to replace the
function $f(k)$, Eq.~(3.6), with $F(k)$, Eq.~(3.41). Then, for
$k_{0}^{-1}<z<1-k_{0}^{-1}$ we obtain \be\label{30v}
\rho_{2}(z)=-F_{0}\,{\rm Re}\biggl(\frac{\e^{-\kappa}}{1-\e^{-2\kappa-2\pi
iz}}\biggr)\,,\ee where
 $F_{0}=F(k_{0}+\Pd)$, $F_{1}=F'(k_{0}+\Pd)$,
 $\kappa=\frac{F_{1}}{2F_{0}}$, and with $z=0$
\bee\nonumber &&\rho_{2}(0)=\Pd I-I_{1}+I_{2},\\&&
I_{1}=\int\limits_{0}^{k_{0}+1/2}{\rm d}k_{z}\, F(k_{z}), \quad I_{2}={\rm
Re}\int\limits_{C_{-}}\frac{{\rm d}k_{z}\,F(k_{z}+k_{0})}{\e^{2\pi i
k_{z}}-1}.\label{31v}\eee Substituting into Eqs.~(3.29) the relations
(3.35), (3.36), (3.38), (3.45), and (3.46), we obtain an expression for
$A(z)$. It looks quite lengthy, although it consists of elementary
functions and simple integrals, numerical calculation of which is easy. It
appears, however, that in the range $k_{0}^{-1}<z<1-k_{0}^{-1}$ a good
approximation for the coefficients $A_{latt}(z)$, $A(z)$ is \be\label{32v}
A_{latt}(z)=2n_{0}\alpha\coth\alpha-\frac{4\pi^{2}}{\omega^{2}}\lambda(z),\quad
A(z)=N_{sat}-\frac{\delta
N_{sat}}{n_{0}}+\frac{2\pi^{2}}{\omega^{2}}\lambda(z),\ee where
$\alpha={\pi}/(\omega_{z}\sqrt{n_{0}})$, \bee \nonumber &&
N_{sat}=\Omega\biggl\{\zeta\biggl(\frac{3}{2}\biggr)+\frac{4}{\sqrt{\pi}}\biggl[\varkappa
\ln(1-\varkappa)+\omega_{x}\ln\biggl(\frac{\omega_{x}}{8\pi\omega_{y}^{2}}
\biggr)\biggr]\biggr\}\,,\quad  \delta
N_{sat}=\frac{\pi\zeta(3)}{2\omega_{y}^{3}\omega_{z}},\\
&&\Omega=\frac{\pi}{8\,\omega_{x}\omega_{y}\omega_{z}}=L_{x}L_{y}L_{z}
\biggl(\frac{m_{mag}k_{B}T}{2\pi\hbar^{2}}\biggr)^{3/2},\quad
\varkappa=\omega_{z}(k_{0}+\Pd).\label{34v}\eee A relative error of the
approximate expression (3.47) in comparison with the exact one,
Eq.~(3.29), when $n_{0}>\omega_{y}^{-2}\gg\omega_{x}^{-2}$, does not
exceed $0.5\times10^{-3}$. As can be seen from Eq.~(3.47), the amplitude
of oscillations $A_{latt}(z)$ decreases with distance along the coordinate
$z$ from boundaries of the film and the background contribution, by
contrast, increases due to the terms proportional to the function
$\lambda(z)$, Eq.~(3.22). The graph of this function for some $n_{0}$ is
shown in Fig. 3.

\begin{figure}[h]
\begin{center}
\includegraphics[height=75mm,keepaspectratio=true]
{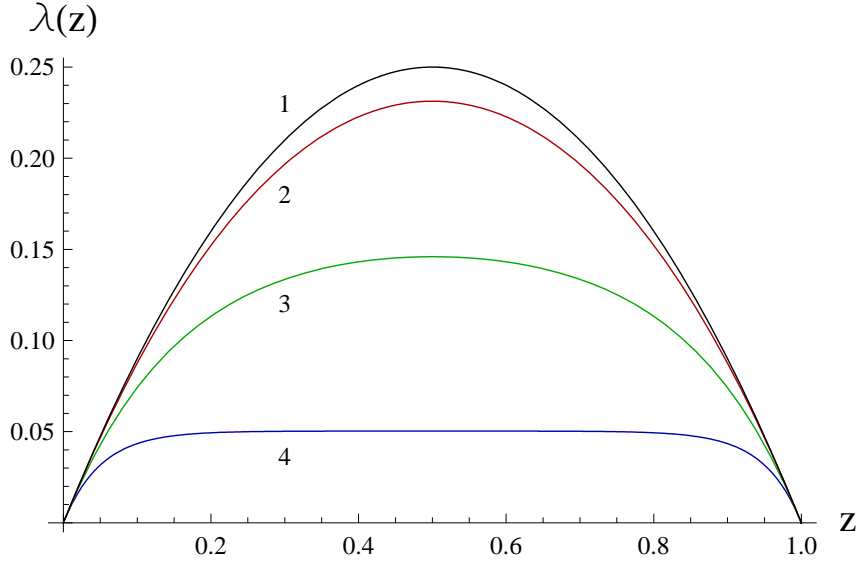} \caption {A behavior of the function $\lambda(z)$ for different
$n_{0}$. The curves 2--\,4 ($j>1$) correspond to
$n_{0}=10^{3-j}\omega_{z}^{2}$, and the curve 1 is for $n_{0}=\infty$.}
\end{center} \end{figure}

Note that the expression (3.47) for $A_{latt}(z)$ in the three-dimensional
case is no different from the one-dimensional case, Eq.~(3.22). However,
the background contribution $A(z)$ is much higher than that of the
one-dimensional problem, Eq.~(3.23), so it is much easier to observe
periodic spatial oscillations of $\rho(z)$ in thin films, and the thinner
the better. Otherwise, an extremely large pumping is required, which would
bring the value $n_0$ to values comparable with $N_{sat}$. It is this
dependence on the pumping level which was observed in Ref. 38.

\section{Conclusion}
Although the conditions for the formation and observation
of the magnon lattice in magnetic films such as YIG are
hard enough, especially at room temperatures, there are a
number of factors that contribute to this effect. In addition to
the decrease of the film thickness noted above, this decrease
of the temperature as well as, what is less obvious, the
increase of the selectivity of the measuring apparatus, which
suppresses frequencies outside the resonance region, to
which in this case the frequency $\nu _0 =\varepsilon _0 /\hbar $ belongs. In YIG it
is equal to $\simeq $2 GHz. In fact, we assume that the frequency
response function of the receiver is characterized by
 \be\label{35v}
\Phi(u_{\qv})=\frac{1}{1+Q^{2}(u_{\qv}-u_{\qv}^{-1})^{2}}\,,\ee where is the $Q$-factor,
 $u_{\qv}=\varepsilon(\qv)/\varepsilon_{0}$.
Then, instead of $N_{sat}$ Eq.~(3.48) there will be observed the other
quantity $N_{obs}$ \be\label{36v}
N_{obs}=\sum_{\qv\neq\qv_{0}}n^{sat}_{\qv}\Phi(u_{\qv}).\ee After
calculations similar to those made in the calculation of Eq.~(3.47), one
can obtain the following approximate expression: \be\label{37v}
 N_{obs}=\Omega\sqrt{\frac{\varepsilon_{0}}{\pi
Q
T}}\biggl[\pi+\arctan(1+2v)-\arctan(1-2v)+\frac{1}{2}\ln\frac{1+2v+2v^{2}}{1-2v+2v^{2}}+
v\ln(1+v^{-4}/4)\biggr],\ee where $v=\varkappa\sqrt{Q T/\varepsilon_{0}}$.
As can be seen from a comparison of Eqs.~(3.48) and (4.3), the suppression
of the background is determined by $\sqrt{\varepsilon_{0}/(Q T)}$. For
example, if $Q=20$ the numerical values of $N_{sat}$ and $N_{obs}$ are
$$N_{sat}\simeq7.42\times10^{16},\quad
N_{obs}\simeq3.68\times10^{14},\quad N_{sat}/N_{obs}\simeq200.$$ The poles
in the complex k-plane of the function Eq.~(4.1) are located far away from
the real axis than the poles of the function $F(k)$, Eq.~(3.28). Therefore
the introduction of the function $\Phi(u_{q})$ does not affect the value
of the coefficient $A_{latt}(z)$, Eq.~(3.47). As a result, when $n_{0}\sim
N_{obs}$ the expression for the magnon density is of very simple form
\bee\nonumber
\bar{\rho}(z)&=&\biggl[2n_{0}\alpha\coth\alpha-\frac{4\pi^{2}}
{\omega^{2}}\lambda(z)\biggr]\cos^{2}\pi k_{0}z+N_{obs}-\frac{\delta
N_{sat}}{n_{0}}+\frac{2\pi^{2}}{\omega^{2}}\lambda(z)\\&\simeq&
2n_{0}\cos^{2}q_{0}l_{z}+N_{obs}\,.\label{38v}\eee Fig. 4 shows graphs of
the spatial density of magnons, calculated according to Eqs.~(4.3) and
(4.4). It is seen that the amplitude of oscillations increases on
increasing the pumping and decreasing the distance from the edge of the
sample along the axis $z$.

\begin{figure}[h]
\begin{center}
\includegraphics[height=32mm,keepaspectratio=true]
{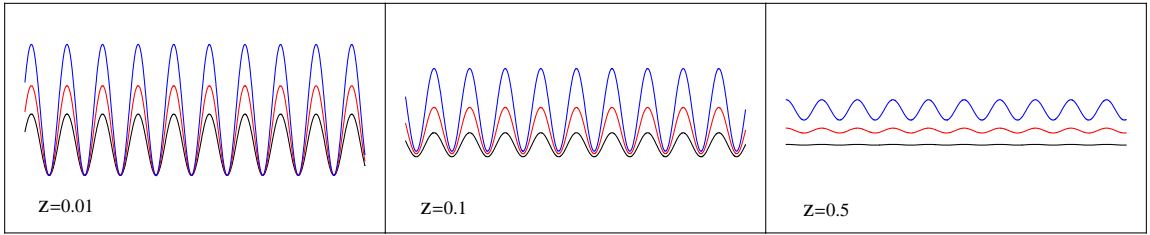} \caption {Oscillation of the spatial density of magnons for
different parts of the film, ${}$ \ \ \ $z=(0.01,\,0.1,\,0.5)$, as a
function of the $n_{0}=(0.5,\,1,\,2)\times 10^{14}\gg n_{0}^{th}$.}
\end{center} \end{figure}

Note that the experiments$^{20}$ can provide the most accurate way to
determine the position of a minimum in the dispersion. The last formula
implies that for pumps that provide the value of $n_0$, which is at least
two orders of magnitude smaller than $N_{sat} $, in a thin ferromagnetic
film one can observe a standing spatial distribution of the density of the
pumped magnons condensed due to relaxation in the lowest state with a
period proportional to $q_0^{-1} $. Such a wave shows up as a lined
(stripe) structure of the magnon density on the axis along which there is
a dip in their dispersion laws $\varepsilon ({\rm {\bf q}})$. A period of
the structure is of $\pi /\tilde {q}_0 $ and in the case of YIG is
$\approx 0.92\ \mu$m, which does not depend on the size of the sample, as
confirmed by measurements.$^{38}$ At the same time, as noted in the same
paper,$^{38}$ the depth of the dips in the formed standing wave of the
spatial distribution of the condensate depends strongly on the intensity
of the pump $I_{pump}$, although it can be shown (see Ref. 21) that the
contrast of the magnon lattice should not depend on the ratio
$I_{pump}/(k_{\scriptscriptstyle B}T)$. Regarding the observations in Ref.
38 of edge dislocations in this lattice, their appearance by pairs
corresponds to the conservation of the topological charge, but the
mechanism of creation of these topological defects requires special
consideration due to the fact that the presence of dislocations in the
structure increases its energy, and hence the periodic condensate with the
extended defects is an excited condensate. In addition, using magnon
lattices one can scatter a light and explore other optical phenomena in
which such a lattice may manifests itself.

\bigskip We acknowledge S. G. Odulov for the discussion of optical
lattices. The work was performed under the program of fundamental research
of the Department of Physics and Astronomy, National Academy of Sciences
of Ukraine.

\section*{Appendix А}
\appendix

As is known, the projections of the spin operator
${\rm {\bf S}}=(S^{x},S^{y},S^{z})$ satisfy the commutation relations$^{35}$
$$\left[ {S^+,S^-} \right]=2S^z, \quad \left[ {S^z,S^\pm } \right]=\pm S^+,
\quad S^\pm =S^x\pm iS^y.\eqno{({\text A.1})}$$

We denote by $\chi _S $ eigenvector of the operator $S^z$ normalized
per unit with the maximum eigenvalue $S$, so that
$$ S^z\chi _S =S\chi _S , \quad S^+\chi _S =0, \quad S^-\chi _S =\sqrt {2S}
\chi _{S-1} , \quad (\chi _S ,\chi _S )=1.\eqno{({\text A.2})}$$
The vector $\chi _{S-1} $ in Eq.~(A.2) is also an eigenvector for
the operator $S^z$ and for it the obvious relations follow from
Eq.~(A.1)
\[
S^z\chi _{S-1} =(S-1)\chi _{S-1} , \quad (\chi _{S-1} ,\chi _{S-1} )=1,
\quad (\chi _{S-1} ,\chi _S )=0.
\]

Let $\left| 0 \right\rangle $ is a ground (vacuum) state of the chain of $L$
spins, oriented along the direction $z$ and with a maximum
value of the $z$-projection, and $\left| l \right\rangle $ is its single-particle excited
state when the $z$-projection of the spin, which is located in
the site $l$, is one less
$$
\left| 0 \right\rangle =\prod\limits_{j=1}^L {\chi _S (j )} , \quad \left|
l \right\rangle =\prod\limits_{j=1}^{.l-1} {\chi _S (j ){\kern 1pt}} \chi
_{S-1} (l)\prod\limits_{j=l+1}^L {\chi _S (j )} , \eqno{({\text A.3})}$$
\[
\langle0|0\rangle =1, \quad \left\langle {0\left| l \right.} \right\rangle
=0, \quad \left\langle {l\left| {l' } \right.} \right\rangle =\delta _{l,
l' } .
\]

If in the spin chain only nearest neighbors interact, its
simplest Hamiltonian can be written in the form $\mathcal{H}=JH$,
where
$$H=S^{-1}\sum\limits_{l=1}^{L-1} {(S^2-{\rm {\bf S}}_l {\rm {\bf S}}_{l-1}
)} , \quad \left[ {{\rm {\bf S}}_{l_1 } ,{\rm {\bf S}}_{l_2 } }
\right]=0.\eqno{({\text A.4})}$$
With the help of Eq.~(A.1) is easy to find that $H\left| 0 \right\rangle =0$; for
all $l$, satisfying the condition $1<l<L$,
$$H\left| l \right\rangle
=2\left| l \right\rangle -\left| {l+1} \right\rangle -\left| {l-1}
\right\rangle,$$
but on the states corresponding to the outermost sites this
Hamiltonian acts differently
$$H\left| 1 \right\rangle =\left| 1 \right\rangle -\left| 2 \right\rangle ,
\quad H\left| L \right\rangle =\left| L \right\rangle -\left| {L-1}
\right\rangle .\eqno{({\text A.5})}$$

Let us introduce a linear combination of single-particle
states
 $$\left| q
\right\rangle =\sum\limits_{l=1}^L {\psi _q (l)} \left| l \right\rangle $$
and act on it by the operator $H$ with taking into account Eq.~(A.5)
$$H\left| q \right\rangle =\sum\limits_{l=1}^L {\left[ {2\psi _q (l)-\psi
_q (l-1)-\psi _q (l+1)} \right]} \left| l \right\rangle ,\eqno{({\text
A.6})}$$
and the wave functions (amplitude) satisfy the free boundary
conditions
$$\psi _q (0)=\psi _q (1), \quad \psi _q (L+1)=\psi _q (L).\eqno{({\text
A.7})}$$
From Eq.~(A.6) it is seen that the bra-vector $\left| q \right\rangle $ is a onemagnon
eigenstate of the spin Hamiltonian
$$H|q\rangle=4\sin ^2(q/2)|q\rangle,$$
if the amplitude $\psi _q (l)$ has the form
 $$\psi _{q} (l )=L^{-\frac{1}{2}}\gamma(q)\cos q (l
-\Pd), \quad q =\frac{\pi k }{L }, \quad k =0,1,...,L -1\,,$$
and its normalization factor is determined from the condition
$\langle q|q'\rangle=\delta _{q, q'}$, using which we find
$$\gamma^{-2}(q)=\frac{1}{L}\sum_{l=1}^{L}\cos^{2}q(l-\Pd)=\Pd(1+\delta_{q,0}).$$

Finally, we note that for commonly used cyclical conditions
to the Hamiltonian Eq.~(A.4) the term $S^{-1}(S^2-{\rm {\bf S}}_L {\rm {\bf S}}_1 )$ is
added, which causes the replacement of the boundary conditions
Eq.~(A.7) by periodic ones (all sites are equivalent)
\[
\psi _q (0)=\psi _q (L), \quad \psi _q (L+1)=\psi _q (1),
\]
As a result, the wave function takes the simple exponential
form
 $$\psi _{q} (l )=L^{-\frac{1}{2}}\gamma(q)\e^{iql}, \quad q
=\frac{2\pi k }{L }, \quad k =0,1,...,L -1\,.$$

\bigskip\noindent
$^1$S. Giorgini, L. P. Pitaevskii, and S. Stringari, Rev. Mod. Phys. 80, 1215
(2008).\\
$^2$T. Giamarchi, C. Ruegg, and O. Chernyshov, Nat. Phys. 4, 198 (2008).\\
$^3$Yu. M. Bunkov, Usp. Fiz. Nauk 180, 884 (2010).\\
$^4$Yu. M. Bunkov and G. E. Volovik, preprint arXiv: 1003.4889 (2012).\\
$^5$A. S. Borovik-Romanov, Yu. M. Bunkov, V. V. Dmitriev, Yu. M.
Mukharskiy,  and\\
$^{}$
D. A. Sergatskov, Phys. Rev. Lett. 62, 1631 (1989).\\
$^6$Yu. D. Kalafati and V. L. Safonov, Sov. Phys. JETP 68, 1162 (1989).\\
$^7$M. I. Kaganov, N. B. Pustylnik, and T. I. Shalaeva, Usp. Fiz. Nauk 167,
197 (1997).\\
$^8$G. A. Melkov, V. L. Safonov, A. Yu. Taranenko, and S. V. Sholom,
J. Magn. Magn. Mater.\\$^{}$ 132, 180 (1994).\\
$^9$S. M. Rezende, Phys. Rev. B 80, 092409 (2009).\\
$^{10}$B. A. Malomed, O. Dzyapko, V. E. Demidov, and S. O. Demokritov,
Phys. Rev. B 81, \\$^{\ \ }$024418 (2010).\\
$^{11}$Yu. M. Bunkov, E. M. Alakshin, R. R. Gazizulin, A. V. Klochkov, V. V.
Kuzmin, \\$^{\ \ }$T. R. Safin, and M. S. Tagirov, Pis’ma v Zh. Eksp. Teor. Fiz. 94,
68 (2011).\\
$^{12}$S. A. Moskalenko and D. W. Snoke, Bose-Einstein Condensation Excitons
and Biexcitons \\$^{\ \ }$(Cambridge University Press, Cambridge 2000).\\
$^{13}$L. V. Butov, A. L. Ivanov, A. Imamoglu, P. W. Littlewood, A. A.
Shashkin, \\$^{\ \ }$V. T. Dolgopolov, K. L. Campman, and A. C. Gossard, Phys.
Rev. Lett. 86, 5608 (2001).\\
$^{14}$V. B. Timofeev, Usp. Fiz. Nauk 175, 315 (2006).\\
$^{15}$Yu. E. Lozovik, A. G. Semenov, and M. Willander, Pisma ZhETF 84, 176
(2006).\\
$^{16}$J. Kasprzak, M. Richard, S. Kundermann, A. Baas, P. Jeambrun, J. M. J.
Kelling, \\$^{\ \ }$F. M. Marchetti, M. H. Szymaska, R. Andre, J. L. Staehli, V.
Savona, P. V. Littlewood, \\$^{\ \ }$B. Deveaud, and L. S. Dang, Nature 443, 409
(2006).\\
$^{17}$R. Balili, V. Hartwell, D. Snoke, L. Pfeiffer, and K. West, Science 316,
1007 (2007).\\
$^{18}$S. O. Demokritov, V. E. Demidov, O. Dzyapko, G. A. Melkov, A. A.
Serga, B. Hillebrands, \\$^{\ \ }$and A. N. Slavin, Nature 443, 430 (2006).\\
$^{19}$A. I. Bugrij and V. M. Loktev, Fiz. Nizk. Temp. 33, 51 (2007) [Low
Temp. Phys. 33, 39 \\$^{\ \ }$(2007)].\\
$^{20}$V. E. Demidov, O. Dzyapko, M. Buchmeier, T. Stockhoff, G. Schmitz, G.
A. Melkov, \\$^{\ \ }$and S. O. Demokritov, Phys. Rev. Lett. 101, 257201 (2008).\\
$^{21}$A. I. Bugrij and V. M. Loktev, Fiz. Nizk. Temp. 34, 1259 (2008) [Low
Temp. Phys. 34, \\$^{\ \ }$992 (2008)].\\
$^{22}$J. Klaers, J. Schmitt, F. Vewinger, and M. Weitz, Nature (London) 468,
545 (2010).\\
$^{23}$J. Klaers, J. Schmitt, T. Damm, D. Dung, F. Vewinger, and M. Weitz,
Proc. SPIE 8600, \\$^{\ \ }$86000L (2013).\\
$^{24}$A. Kruchkov and Yu. Slyusarenko, Phys. Rev. A 88, 013615 (2013).\\
$^{25}$F. Li, W. M. Saslow, and V. L. Pokrovsky, Sci. Rep. 3, 1372 (2013).\\
$^{26}$M. H. Anderson, J. N. Ensher, M. R. Matthews, C. E. Wieman, and E. A.
Cornell, \\$^{\ \ }$Science 269, 198 (1995).\\
$^{27}$C. C. Bradley, C. A. Sackett, J. J. Tollett, and R. G. Hulet, Phys. Rev.
Lett. 75, \\$^{\ \ }$1687 (1995).\\
$^{28}$K. B. Davis, M.-O. Mewes, M. R. Andrews, N. J. van Druten, D. S.
Durfee, D. M. Kurn, \\$^{\ \ }$and W. Ketterle, Phys. Rev. Lett. 75, 3969 (1995).\\
$^{29}$L. D. Landau and E. M. Lifshitz, Statistical Physics (Pergamon Press,
Oxford, 1980), \\$^{\ \ }$p. 1.\\
$^{30}$C. Kittel, Quantum Theory of Solids (John Wiley and Sons, Inc., NewYork
1963).\\
$^{31}$N. B. Brandt and V. A. Kul’bachinskii, Quasiparticles in Condensed
Matter \\$^{\ \ }$(Fizmatlit, Moscow, 2005).\\
$^{32}$B. A. Kalinikos and A. N. Slavin, J. Phys. C 19, 7013 (1986).\\
$^{33}$Yu. A. Izyumov and R. P. Ozerov, Magnetic Neutronography (Plenum
Press, \\$^{\ \ }$London 1970).\\
$^{34}$V. L. Vinetskii, N. V. Kukhtarev, S. G. Odulov, and M. S. Soskin, Sov.
Phys. Uspekhi 22, \\$^{\ \ }$742 (1979).\\
$^{35}$A. I. Akhiezer, V. G. Bar’yakhtar, and S. V. Peletminskii, Spin Waves
\\$^{\ \ }$(North Holland, Amsterdam, 1968).\\
$^{36}$A. S. Davydov, Teoriya Tverdogo Tela (Theory of Solid State), (Nauka,
Moscow, 1976).\\
$^{37}$A. P. Prudnikov, Yu. A. Brychkov, and O. I. Marichev, Integrali i Ryady
\\$^{\ \ }$(Integrals and Series), (Fizmatlit, Moscow, 2003).\\
$^{38}$P. Nowik-Boltyk, O. Dzyapko, V. E. Demidov, N. G. Berloff, and O.
Demokritov, \\$^{\ \ }$Nat. Sci. Rep. 2, 482 (2012).

\end{document}